\documentclass[conference]{IEEEtran}
\IEEEoverridecommandlockouts

\usepackage{cite}
\usepackage{amsmath,amssymb,amsfonts}
\usepackage{algorithmic}
\usepackage{multirow}
\usepackage{graphicx}
\usepackage{textcomp}
\usepackage{xcolor}
\usepackage{orcidlink}
\usepackage{url}
\usepackage{csquotes}
\usepackage{booktabs}
\def\BibTeX{{\rm B\kern-.05em{\sc i\kern-.025em b}\kern-.08em
    T\kern-.1667em\lower.7ex\hbox{E}\kern-.125emX}}

\begin{document}

\title{The Influence of HEXACO Personality Traits on the Teamwork Quality in Software Teams - A Preliminary Research Approach\\
}

\author{
\IEEEauthorblockN{Philipp M. Zähl \orcidlink{0000-0003-3302-4415} }
\IEEEauthorblockA{\textit{Institute of Digitalization Aachen} \\
\textit{FH Aachen University of Applied Sciences}\\
Aachen, Germany \\
zaehl@fh-aachen.de}
\and
\IEEEauthorblockN{Sabine Theis \orcidlink{0000-0002-3422-3734}}
\IEEEauthorblockA{\textit{Institute of Software Engineering} \\
\textit{German Aerospace Center (DLR)}\\
Cologne, Germany \\
sabine.theis@dlr.de}
\and
\IEEEauthorblockN{Martin R. Wolf \orcidlink{0009-0008-2227-5116}}
\IEEEauthorblockA{\textit{Institute of Digitalization Aachen} \\
\textit{FH Aachen University of Applied Sciences}\\
Aachen, Germany \\
m.wolf@fh-aachen.de}
}

\maketitle

\begin{abstract}
Although software engineering research has focused on optimizing processes and technology, there is a growing recognition that human factors, particularly teamwork, also significantly impact optimization. Recent research suggests that developer personality has a strong influence on teamwork. In fact, personality considerations may have a greater impact on software development than processes and tools. This paper aims to design a study that measures the impact of HEXACO personality traits on the Teamwork Quality (TWQ) of software teams. A preliminary data collection (n=54) was conducted for this purpose. The analysis showed that several personality traits, as well as their composition, had a significant impact on TWQ. Additionally, other variables, such as the proportion of women and age distribution, also affected TWQ. The study's initial results demonstrate the usefulness and validity of the study design. The results also suggest several opportunities to improve teamwork in IT organizations and avenues for further research.
\end{abstract}


\section{Introduction} \label{sec_introduction}
IT systems and the technologies used in them are becoming larger and more complex. This not only increases the need to develop software in teams, but also to optimize software development in terms of efficiency. \cite{zahl_teamwork_2023, dutra_what_2015} The shortage of skilled workers and the increasing demand for shorter working hours could reinforce this need \cite{lpbnrw_fachkraftemangel_2023, rieger_vier-tage-woche_2023}. Research has traditionally focused on optimizing processes and improving technology. However, there is now a growing recognition that human factors, particularly teamwork, also play a crucial role in optimization. Developer personality, in particular, has a significant impact on teamwork, making it an increasingly important area of research. Current research findings also support the assumption that personality has a greater influence on software development than processes and technologies. \cite{cruz_forty_2015, felipe_psychometric_2022} Much of personality research in the software engineering environment uses the \textit{Myers-Briggs Type Indicator} (MBTI) from 1944 as its personality model \cite{cruz_forty_2015, felipe_psychometric_2022}. 
However, the MBTI is not considered to be an accurate description due to its low reliability, dichotomous ("either-or") preferences, and the Barnum effect in subjects' self-assessments as an unreliable model of personality \cite{russo_anecdote_2022, boyle_myers-briggs_1995, pittenger_measuring_1993}. In contrast, the use of more recent personality models promises more meaningful results \cite{russo_anecdote_2022, de_vries_zes_2009}.

The aim of this paper is to develop a study design for the research question (RQ) \enquote{What influence do personality traits have on the quality of teamwork in german software teams?} and to evaluate it on a first sample. For this purpose, Section \ref{sec_background} introduces the most important terms and provides an insight into the current state of research. Section \ref{sec:hypotheses} presents the hypotheses relevant to the RQ, which will be tested using the methodology described in section \ref{sec_method}. The results of the preliminary study are then presented, interpreted, and discussed in Section \ref{sec:results}. The limitations of the study design are also critically examined. Finally, further research possibilities, implications for further RQ-related research, and first possible implications for practice are presented.

\section{Related Work} \label{sec_background}

\subsection{Personality}

Over the years, different perspectives have emerged in personality research, such as the dispositional, biological, psychoanalytic, or learning theory perspective \cite{scheier_model_1988}. The dispositional perspective includes the trait and type theory presented later in this paper, which is one of the most commonly used theories in organizational psychology and in personality research in software development. This paper focuses on this dispositional perspective. \cite{cruz_forty_2015} Therefore, the following personality definition will be applied: Personality, according to the trait-theoretic or factor-analytic approach, is a pattern of behavior or combination of traits of temperament, emotions, intellect, and ways of acting, communicating, and moving that persists over time. Personality can thus also be described as the \textit{normal/ordinary} behavior of a person. \cite{apa_personality_2023, roth_personlichkeit_2011, becker_grundlagen_2021} According to this approach, personality is described in terms of various traits (or also: characteristics, traits, dispositions, factors, dimensions). Their influence is seen as a mediator between stimulus and reaction \cite{bunge_philosophie_1990, becker_grundlagen_2021}. However, it should be noted that like many other models and theories in psychology, there are different definitions (see for example \cite{guilford_guilford-zimmerman_1949, eysenk_scientific_1953, allport_europaische_1959, cattell_personality_1973, ryckman_theories_2004}). 

Approximately 20~\% of the personality is shaped by the social environment. This social environment can differ according to gender: Children of certain age groups prefer a same-sex circle of friends \cite{martin_stability_2001} and also upbringing and education can partly differ according to gender \cite{buchmann_gender_2008}. In this respect, the question arises to what extent personality is related to sociocultural (learned) gender roles. Gender roles or stereotypes have indeed been questioned since the women's movement in the 1970s, especially with regard to power structures and economic dependencies. Nevertheless, traditional role models are still present today, such as the supposed \enquote{Boys don't cry} principle that men must not show any feelings in order not to be considered \textit{weak}, or that women who are not at home with their children due to their jobs would not love them (cf. \textit{Stereotype Content Model}). \cite{vogel_boys_2011, wolter_aufwachsen_2020} That these role models have persisted to this day is partly because people tend to equate a person's activities with his or her personality traits. Thus, for example, if mothers predominantly stay at home to care for the children, mothers or women are generally thought to have a more caring personality. On the other hand, gender roles are also modeled in part by parents. And since children imitate their parents' behavior and in turn influence other children, the gender stereotypes of the previous generation also persist. In both cases, a person will tend to conform his or her behavior to the gender role model in order to conform to the social norm and avoid social consequences. The \textit{self-fulfilling prophecy} can also be applied to gender stereotypes. \cite{wolter_aufwachsen_2020}
In summary, gender roles have a high impact on personality. Accordingly, gender observations and expressions should not be viewed in this paper as solely the result of biological sex. 

Personality models are a simplified representation of personality. For this purpose, personality is decomposed into individual traits according to the definition of \cite{guilford_guilford-zimmerman_1949} and \cite{eysenk_measurement_1976}. The features of a model can be either personality types or personality traits (trait approach/theory). Personality \textbf{type} (or profile) models contain discrete (dichotomous) traits, meaning they do not allow for gradation: \enquote{When a person is assigned to a type, he or she cannot belong to any other type from the same classification system. Many people like to use personality types in everyday life because they help simplify the complicated process of understanding other people}. \cite{gerrig_psychologie_2008, becker_grundlagen_2021} Personality types are helpful in constructing a comparable ideal personality. At the same time, they promote \textit{pigeonhole thinking} and are unrealistic due to the restriction to a few characteristics as well as a rudimentary personality analysis. \cite{jumpertz_personlichkeitstypologien_2004, hossiep_personalauswahl_2015, becker_grundlagen_2021} Personality \textbf{trait} models, on the other hand, allow for gradation, assuming that individuals cannot be completely assigned to a single type. Therefore, each trait is represented by a number or percentage. According to Bonini's paradox, trait-theoretic models of personality are more difficult to understand. Trait theory also assumes that personality traits are stable over time and across situations. Although people do not behave the same way in all situations, there is an assumed consistency of behavior and experience across many situations. \cite{rammsayer_differentielle_2016, becker_grundlagen_2021}

In this paper, personality will be described according to the trait theory approach. The most common operationalization of personality in psychology according to this approach is the \textit{Big Five Inventory} (BFI)\footnote{Or also: Big Five, Five-Factor-Model (FFM), OCEAN Model (acronym from the traits), NEO-PI (\textit{NEO} = acronym from the traits Neuroticism, Extraversion \& Openness | \textit{PI} = Personality Inventory), NEO-PI-R (\textit{R} = \textit{Revised}, revised version of the NEO-PI)}. The BFI was created in 1985 to replace the MBTI personality model, which was in use at the time. Even then, it was believed that either Jung's theory was incorrect or the MBTI was incorrectly operationalized and therefore inadequate for personality assessment. \cite{mccrae_reinterpreting_1989} The BFI, on the other hand, is a lexical model of personality and is based on the assumption that all personality traits can be described by one adjective. To develop the BFI, all personality-describing words were collected and reduced to five independent, culturally stable factors using factor analysis.
The most significant development of the BFI today can be traced back to \cite{costa_neo_1985} or the corrected version of \cite{costa_neo_1992} and is a standard model in personality psychology. This study uses the newer, but less well-known, HEXACO model instead. The HEXACO model was developed in an attempt to transfer the methodology of the original English-language BFI to other languages. In the process, a sixth dimension, Honesty-Humility, was identified. \cite{ashton_six-factor_2004} HEXACO can therefore be seen as a supplement to the BFI and can be represented as shown in Table \ref{tab:hexaco_dims_facets} \cite{iller_handbuch_2021, ashton_empirical_2007, russo_anecdote_2022}.

\begin{table}[htb]
\centering
\caption{HEXACO dimensions and facets}
\label{tab:hexaco_dims_facets}
\begin{tabular}{llp{4.8cm}}
\multicolumn{2}{l}{\textbf{Dimensions}} & \textbf{Facets}                                                            \\
\textbf{H}   & Honesty-humility         & sincerity, fairness, material frugality, modesty                           \\
\textbf{E}   & Emotionality             & anxiousness, concern, dependence, sentimentality                           \\
\textbf{X}   & Extraversion             & social sense of self, social courage, sociability, vivacity                \\
\textbf{A}   & Agreeableness            & willingness to forgive, forbearance, willingness to compromise, gentleness \\
\textbf{C}   & Conscientiousness        & organization, diligence, perfectionism, prudence                           \\
\textbf{O}   & Openness to experience   & aesthetic appreciation, inquisitiveness, creativity, unconventionality    
\end{tabular}
\end{table}

Several studies can demonstrate that the new Honesty-Humility dimension is clearly determinable and a good complement for the BFI \cite{ashton_hexaco_2014, saucier_recurrent_2009, russo_anecdote_2022}. The authors of \cite{ashton_empirical_2007} argue that HEXACO depicts personality in a more sophisticated and comprehensive way than the BFI. Further, they see better interpretability because the HEXACO facets correspond to \enquote{naturally observable traits}, which also means that they are more informative for practice. The requirements for linguistic and cultural stability of HEXACO are also better met due to the optimized model concept. Meanwhile, HEXACO is becoming a new standard in social psychology \cite{ashton_hexaco_2014}. Finally, the HEXACO model has rarely been used in software engineering research, and the honesty-humility dimension has not yet been applied in this context. Thus, the impact of moral traits (e.g., honesty, humility, altruism, and justice) on SE remains unexplored. Overall, applying the HEXACO model to personality assessments of software developers is considered to have great research potential.

\subsection{Teamwork}

In this work the definition of software teams of \cite{zahl_teamwork_2023} will be adopted. According to this definition software teams are groups of two or more persons, who work together on one or more goals in the range of the software engineering. Thus not only software development teams fall into the consideration, but also teams e.g. in the ranges IT-security or IT-architecture. Teams and especially software teams can be classified as dynamic (word origin: Greek. \textit{dynamikós}: wealthy, effective, powerful, strong - \cite{dwds_digitales_2023}) systems \cite{zahl_teamwork_2023}. A dynamic system is one that changes or evolves over time. This can occur in a variety of ways, including changes in the size or structure of the system, changes in the relationships among the elements of the system, or changes in the elements themselves. \cite{tuckman_stages_1977} categorize the processes of change within teams by the phases of \textit{forming}, \textit{storming}, \textit{norming}, and \textit{performing}, which are cycled through.
Team performance typically changes over the course of this cycle \cite{vater_kartenset_2021, tuckman_stages_1977}. The duration of the phases can vary depending on the team and the general conditions and can therefore not be scheduled across the board. Concept studies do exist, such as \cite{nguyen_predictive_2016}, in which development phases are assigned to calendar weeks of a university semester. However, artificially shortening the development phases in this way can have a negative impact on teamwork in subsequent phases and should therefore be avoided \cite{adelakun_it_2003}.

\begin{figure}[ht]
\centering
\includegraphics[width=0.7\columnwidth,trim={4.8cm 2cm 8cm 2cm},clip]{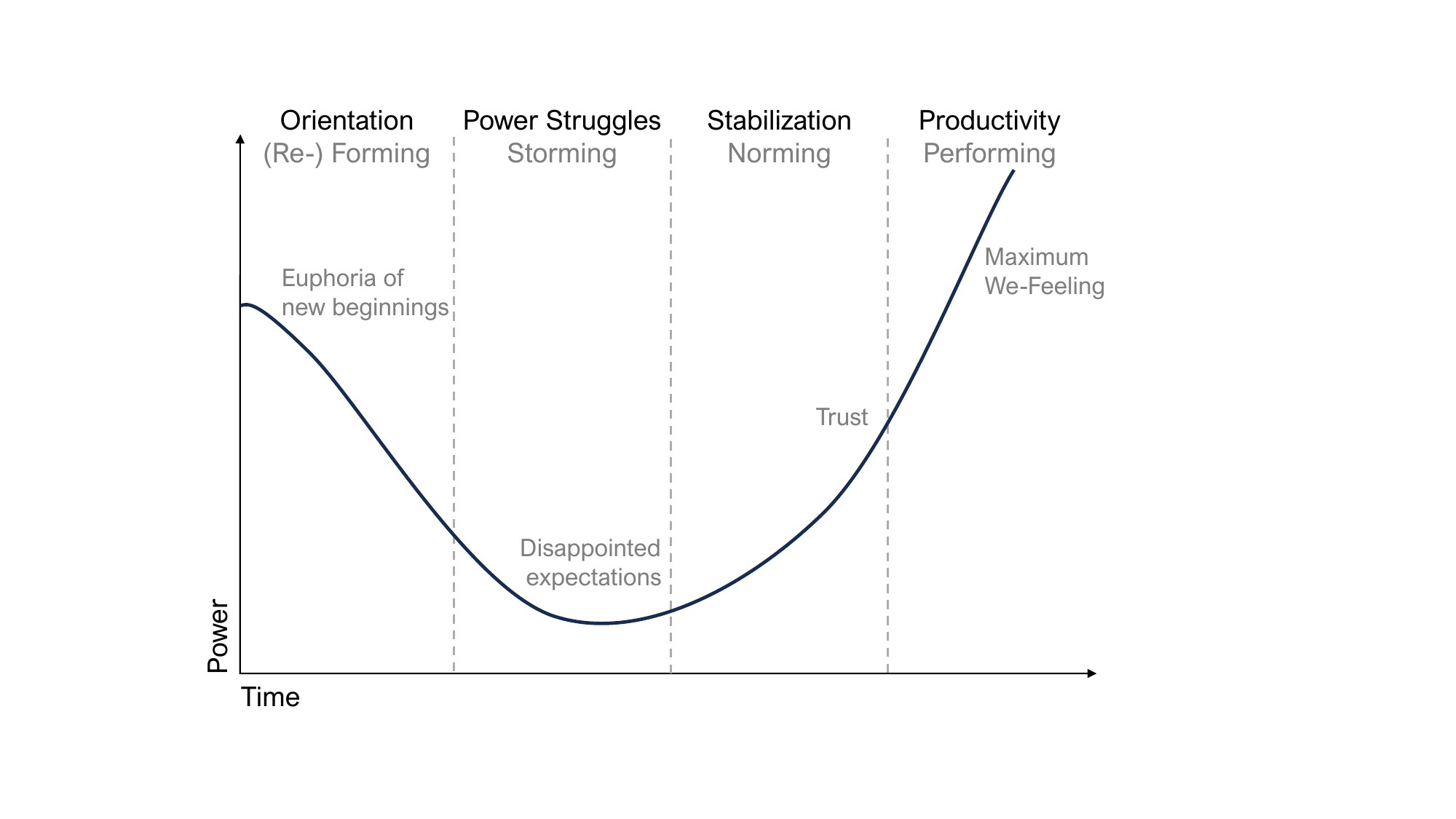}
\caption{Team performance in the team development phases according to \cite{tuckman_stages_1977} from \cite{vater_kartenset_2021}}
\label{img_team_leistungskurve}
\end{figure}

Various team metrics exist for evaluating teams. One of them is Teamwork Quality (TWQ), that was defined by \cite{hoegl_teamwork_2001}. TWQ consists of the facets of communication (CMNC), coordination (CRDN), balance of member contributions (BLNC), mutual support (SPRT), commitment (COMT) and cohesion (COHS). It is based on the findings on the quality of work groups according to \cite{hackman_design_1987}. Hackman defined TWQ as \enquote{the interaction of group members that contributes to achieving a common goal}. He noted that TWQ is influenced by several factors, such as clarity of group goals, work structure, and group composition. Thus, TWQ should be understood as a snapshot, as it can vary depending on the current task or when changes occur in the team. \cite{singh_longitudinal_2022, hackman_design_1987, hoegl_teamwork_2001} Nevertheless, TWQ has a significant impact on team performance and personal success of team members \cite{hoegl_teamwork_2001, purna_sudhakar_soft_2011}.

\section{Hypotheses} \label{sec:hypotheses}
Based on the RQ and the presented state of research, the hypotheses to be tested in this paper are formulated below. The aim is to establish a possible relationship between the variables personality and TWQ and their subordinate facets.

\begin{enumerate}
    \item[H1] A personal TWQ estimation correlates with certain personality traits. \\
    \textit{Background: Each team member evaluates the TWQ of his team (individual evaluation) according to his subjective perception. For this reason, a team's TWQ can vary from team member to team member, depending on their individual perspectives.
    }
    \item[H2] Team TWQ correlates with certain personality traits. \\
    \textit{Background: Individual scores must be consolidated, e.g. by statistical averaging, to provide a tolerably objective team TWQ score. It is assumed that TWQ total scores are related to individual personality traits.}
    \item[H3.1] Team TWQ correlates with certain personality traits of the average team member.\\
    \textit{Background: Teams can be seen as a set of different personalities. To simplify the description of this set, it is computationally possible to average the personality of the team (team personality).
    }
    \item[H3.2] The TWQ of teams correlates with the diversity of certain personality traits within the team.\\
    \textit{Background: The diversity of teams influences teamwork and team performance depending on the trait in question \cite{liang_effect_2007, zahl_teamwork_2023}. Therefore, for this hypothesis, we will examine which personality traits have a positive influence on TWQ when they are differentially expressed within the team.}
\end{enumerate}

Two levels of perspective can be identified from the aforementioned hypotheses: The individual level and the team level. While the individual level depicts the characteristics of a single person, these characteristics must be consolidated at the team level. A classification of the hypotheses in the level scheme is shown in table \ref{tab_hypotheses_matrix}.

\begin{table}[htb]
\centering
\caption{Classification of the hypotheses in the observation levels}
\label{tab_hypotheses_matrix}
\begin{tabular}{ll|ccc}
 & \textbf{Personality} & \multicolumn{1}{c}{\textbf{-}} & \multicolumn{1}{c}{\textbf{Indiv.}} & \textbf{Team} \\
\textbf{TWQ} &  & \multicolumn{1}{l}{} & \multicolumn{1}{l}{} & \multicolumn{1}{l}{} \\ \hline
\multicolumn{1}{c}{\textbf{-}} &  & \multicolumn{1}{c}{-} & \multicolumn{1}{c}{Not relevant} & \begin{tabular}[c]{@{}c@{}}Descriptive \end{tabular} \\ 
\textbf{Indiv.} &  & Not relevant & \multicolumn{1}{c}{H1} & \begin{tabular}[c]{@{}c@{}}Not relevant\end{tabular} \\ 
\textbf{Team} &  & \begin{tabular}[c]{@{}c@{}}Descriptive \end{tabular} & H2 & H3.x
\end{tabular}
\end{table}

In the further course of this paper, these observation levels will be referenced more frequently. The naming scheme \enquote{$x$-$y$} is used, where $x$ represents the personality axis and $y$ the TWQ axis.

\section{Research Design} \label{sec_method}


The target population to be studied comprises German software developers or software development teams. To simplify data collection, the target population is reduced to the subpopulation of student software teams. These student software teams should consist of at least three people and be active in one software engineering task area. Strictly speaking, student software developers do not fit the definition of software professionals, as they do not have a university degree and usually do not have sufficient work experience. 
Nevertheless, the investigation of this subpopulation will initially be sufficient to verify the study design and identify preliminary data trends.

\begin{figure}[ht]
\centering
\includegraphics[width=0.6\columnwidth,trim={3.4cm 3cm 3.4cm 1.5cm},clip]{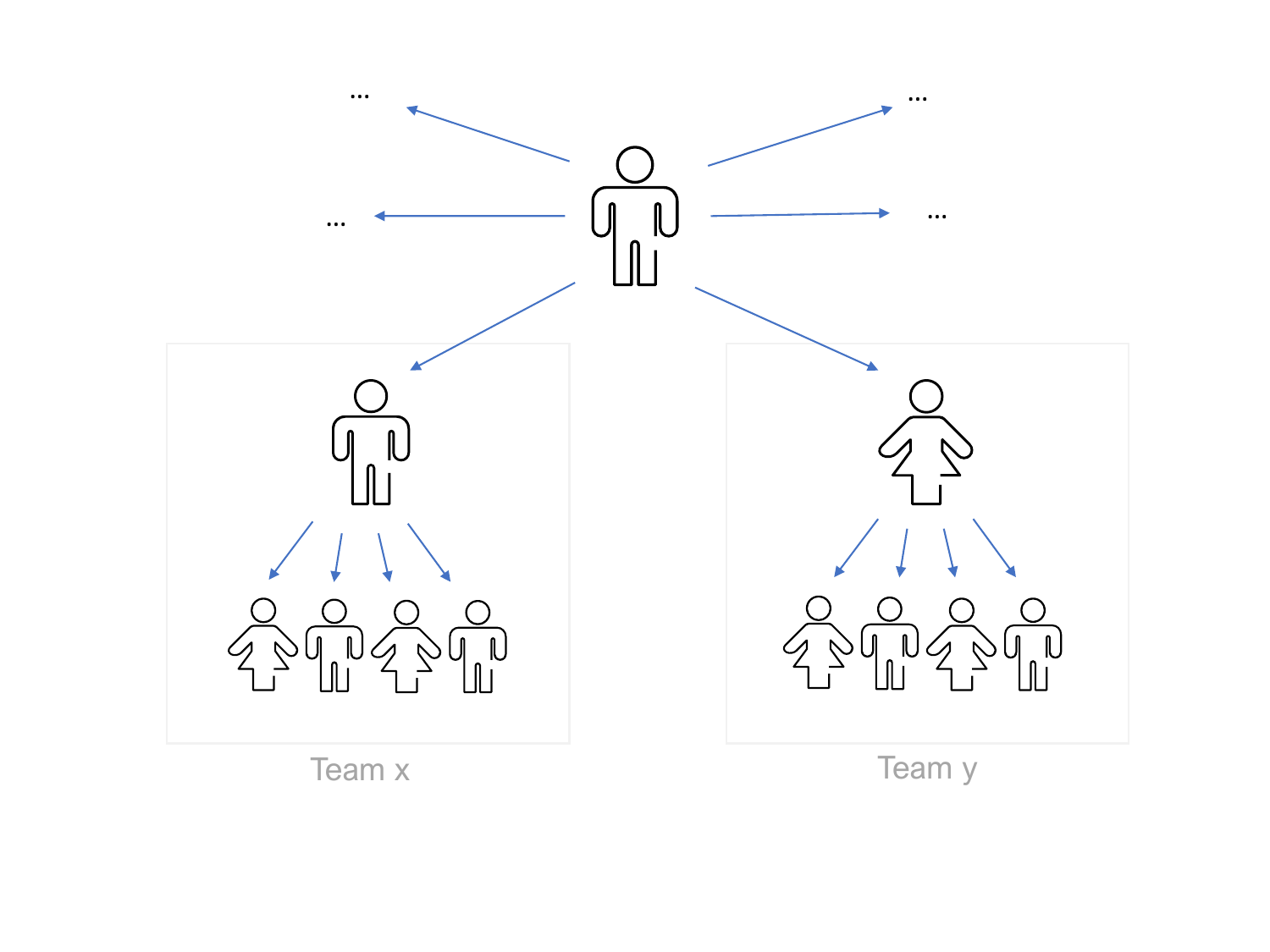}
\caption{Used sampling method}
\label{img_sampling}
\end{figure}

A sample study was conducted to quantitatively investigate the influence of developer personality on TWQ. Sampling studies are the most common research strategy in software engineering research and have the greatest potential for generalizability. Non-probability sampling will be applied for this study. This means that the sample is chosen to be non-random, which makes it faster and less expensive to conduct. \cite{stol_abc_2018} Additionally, the chosen sampling method is \enquote{Snowball-Sample} because the study requires entire teams rather than individuals (see Figure \ref{img_sampling}). Team leaders, who supervise teams meeting the above conditions, are contacted via email, LinkedIn chat, or verbally. They can then motivate their team to participate. This approach enables us to identify the participating teams beforehand and provide a suitable team selection in the questionnaire. Additionally, snowballing streamlines canvassing, as only one person (i.e., the team leader) needs to be identified per team.
A quantitative survey using a questionnaire was chosen as the research method for data collection. 
This questionnaire is divided into three parts: The first part includes the HEXACO personality test, while the second part captures the TWQ assessment. The third part asks for additional demographic data. The HEXACO-PI-R questionnaire operationalizes personality according to the HEXACO personality model \cite{howard_discriminant_2020}. It consists of 60 items that query a number of personality facets using two to four self-assessment items. Grouped and statistically averaged, this yields the associated personality trait. It is taken into account that HEXACO, respectively the BFI in particular, was originally developed in English. For this work, however, the German HEXACO-PI-R translation is used, which, as well as those of other languages, was equally positively evaluated \cite{thielmann_hexaco-100_2020, moshagen_faktorenstruktur_2014}. The questionnaire has been partially adapted to the update of \cite{iller_handbuch_2021} to reflect contemporary expressions and gender-sensitive language.
The TWQ is operationalized by means of the questionnaire according to \cite{hoegl_teamwork_2001} containing six facets with a total of 37 items. The resulting TWQ is also subjective and cannot be used alone to evaluate the entire team, since this individual TWQ only represents the subjective perception of a team member. Thus, the TWQ rating may vary among the team members.
The final questionnaire has been referenced in chapter \ref{sec_suppleMat}. A correlation analysis was performed to analyze the data and identify possible correlations between personality and TWQ. These correlations are then tested for significance using T-test and Cronbach's alpha. Previously, all incomplete participations were removed from the data set. In addition, all information that was no longer required was removed or pseudonymized.

Due to the sensitivity of the topic, especially when surveying students and collecting sensitive data such as personality in the context of studies and work, ethical standards should be particularly observed. The privacy and well-being of the participants were always respected throughout the research process. The students participated voluntarily. Care was taken to ensure that participants had the freedom to accept or decline their participation in order to preserve their autonomy. When using the anonymous personality and TWQ test, care was taken to protect sensitive personal information and maintain the confidentiality of the participants. The team name was only required for clear assignment to the teams, which meant that this information could be pseudonymized once data collection was complete. It is not possible to draw conclusions about the teams at a later date. The results were analyzed while maintaining the anonymity of the participants and the confidentiality of sensitive information. The separation of genders in the analysis was not intended to highlight one gender, but to obtain reliable and generalizable results for all genders. The correlations revealed between personality traits and TWQ scores are not intended to characterize certain personality trait characteristics as more or less important (see \cite{costa_neo_1985}).

\section{Results} \label{sec:results}

We conducted the study in 2023. A total of 54 individuals participated, of which 42 are man ($77~\%$) and 12 are women ($22~\%$). The youngest participant is 18 years old and the oldest participant 39 years old, giving a range of 21 years and an average age of $24.4$ years. Most participants ($72~\%$) are in a bachelor's degree program in (business) computer science or its equivalent. Eight participants are pursuing a master's degree in (business) computer science or similar. The participants could be assigned to 13 different real teams working together, of which eight ($61~\%$) meet the specified minimum number of three persons. These teams reach a combined total of 47 people, so on average the approved teams have a participant count of $5.8$~people (median $4$). The team with the most survey participants includes 12 people. Two teams are student teams from companies, two teams are university-based development/research projects, and the remaining are unpaid university projects completed to earn credit points (ECTS). Eight of the twelve teams are located at University of Applied Sciences Aachen. The HEXACO results of all participants are shown in Table \ref{tab:hexacoresults}. According to this, C is the highest and E is the lowest. Furthermore, the dimensions E and X have the highest range.

\begin{table}[htbp]
\caption{\label{tab:hexacoresults}HEXACO Results}
\centering
\begin{tabular}[t]{ r r | c c c }
\hline
~ & ~ & Total & Male & Female\\
\hline
~ & n & 54 & 42 & 12 \\
\hline
\multirow{6}{*}{\rotatebox[origin=c]{90}{Mean}} & H & 3.292 & 3.198 & 3.618 \\
~ & E & 2.936 & 2.765 & 3.535 \\
~ & X & 3.133 & 3.188 & 2.941 \\
~ & A & 3.241 & 3.2 & 3.382 \\
~ & C & 3.407 & 3.318 & 3.719 \\
~ & O & 3.198 & 3.185 & 3.247 \\
\hline
\multirow{6}{*}{\rotatebox[origin=c]{90}{SD}} & H & ~ & 0.694 & 0.51 \\
~ & E & ~ & 0.765 & 0.743 \\
~ & X & ~ & 0.692 & 0.851 \\
~ & A & ~ & 0.548 & 0.482 \\
~ & C & ~ & 0.498 & 0.232 \\
~ & O & ~ & 0.585 & 0.527 \\
\hline
\end{tabular}
\end{table}

Female participants (see Table \ref{tab:hexacoresults}) have significantly higher scores than male participants in E ($+43~\%$), H ($+19~\%$), and C ($+17~\%$), slightly higher scores in A ($+8~\%$) and O ($+3~\%$), and a lower score in X ($-11~\%$). These results are also largely consistent with the study results according to \cite{russo_gender_2022}. At the same time, female participants have the highest dispersion in X ($SD = 0.851$) and E ($SD = 0.753$), while male participants have the highest dispersion in E ($SD = 0.765$) and H ($SD = 0.694$).
On average, participants rate TWQ as $3.708$ ($67.705~\%$), with TWQ facets taking the following values: CMNC $3.71$ ($67.75~\%$), CRDN $3.63$ ($65.75~\%$), BLNC $3.75$ ($68.75~\%$), SPRT $3.98$ ($74.5~\%$), COMT $3.47$ ($61.75~\%$) and COHS $3.67$ ($66.75~\%$). Accordingly, SPRT is rated the highest, while COMT is rated the lowest. The greatest dispersion is seen in CRDN. Female participants rate all TWQ facets equally or higher than male participants. In particular, COHS ($+14~\%$), SPRT ($+7~\%$), and thus TWQ overall ($+5~\%$) are perceived more positively.
As already announced in chapter \ref{sec_method}, the collected data are compared in the context of correlation analyses. This is done according to the level representation from table \ref{tab_hypotheses_matrix}. The data tables have been outsourced for improved reading flow and can be viewed via the reference in chapter \ref{sec_suppleMat}.


The HEXACO dimensions reach the alpha values $\alpha_H=0.75$, $\alpha_E=0.85$, $\alpha_X=0.81$, $\alpha_A=0.67$, $\alpha_C=0.8$, and $\alpha_O=0.6$. Thus, H, E, X, and C score acceptable to good, while A and O score too low. With an average value of $0.74$, the HEXACO dimensions nevertheless achieve an acceptable result overall.
Compared to the alpha values of the original, English HEXACO-60 according to \cite{ashton_hexaco-60_2009}, where all dimensions have a value between $0.73$ and $0.80$, partially comparable values could be achieved in this study. 
A similar comparison is made with the alpha values of the German translation of HEXACO according to \cite{moshagen_faktorenstruktur_2014}, where values between $0.74$ and $0.83$ were obtained. The updated HEXACO-PI-R1 questionnaire according to \cite{iller_handbuch_2021} reaches alpha values between $0.88$ and $0.94$ - a sample size is not mentioned.\\
The negative deviation of A and O can possibly be attributed to the comparatively small sample size. In the case of openness, it is also conceivable that IT students would give partially contradictory answers. For example, an IT student could have little interest in creative work and at the same time a high interest in music or history, which would have a negative effect on their alpha value in the applied operationalization of the openness dimension.
The TWQ facets reach the following $\alpha$ values: $\alpha_{\text{CMNC}} =0.82$, $\alpha_{\text{CRDN}} =0.7$, $\alpha_{\text{BLNC}} =0.65$, $\alpha_{\text{SPRT}} =0.92$, $\alpha_{\text{COMT}} =0.63$, $\alpha_{\text{COHS}} =0.89$.
CMNC, CRDN, SPRT, and COHS thus score acceptable to excellent, while BLNC and COMT score too low.

\subsection{Comparison}

In this section, the HEXACO values collected will be compared with those of some other studies as an example. The studies considered are: Croatian students from \cite{babarovic_hexaco_2013}, Canadian students from \cite{ashton_hexaco-60_2009}, Japanese psychology students from \cite{wakabayashi_sixth_2014} and International software developers from \cite{russo_anecdote_2022}. Since the HEXACO model has hardly been used in studies in the field of software engineering so far, the respondents in some samples are located outside of computer science. Therefore, it is not possible to validate the figures presented in this paper by comparing them with the other study results, since the groups surveyed are only comparable to a limited extent. Instead, however, an initial statement can be made as to whether gender-related differences in HEXACO scores have also been recorded in other studies and whether personality characteristics can be identified in comparison with other studies.

Obviously, the women interviewed in this sample are more introverted than the women in the other studies. Also, this study is the only one in which the men are more extroverted than the women. The reverse is true for agreeableness, as the other studies show higher agreeableness among men, while this study finds higher agreeableness among women.
The group that appears most similar to this study is that of international software developers from \cite{russo_anecdote_2022}. Despite this perceived similarity, some salient differences can be identified: The international software developers scored much higher in honesty ($+15~\%$ to $+18~\%$) and openness ($+27~\%$ to $+36~\%$), while significantly lower scores were measured in agreeableness ($-18~\%$ to $-25~\%$). The differences could be due to differences in culture and educational system, as the proportion of German respondents is only $2.5~\%$. British and Americans, on the other hand, account for a combined $57.2~\%$, so potential cultural influence would trivially be strong. In this case, however, the cultural stability of the HEXACO variables would be in doubt. Last, the selection procedures used to obtain the sample and the composition of the participants differ.

\section{Discussion}

\subsection{Interpretation}

The two genders show partly strongly different values in the HEXACO dimensions. It is assumed that these differences are due to genetic-biological influence as well as cultural imprints. Among men, there is a wide variation in emotionality. Social stigma may have led some men to try to withhold their emotions in order to conform to the gender norm. \cite{vogel_boys_2011, wolter_aufwachsen_2020, eagly_gender_1984, meuser_geschlecht_2010} A similar explanation would be possible for the high dispersion in emotionality among women.

In men, the individual assessment of TWQ depends significantly on the three personality dimensions E (-), X and A. The negative correlation with E could be due to the fact that emotional men might react more sensitively to interpersonal conflicts or tensions in the team. It would also be conceivable that emotional men tend to suppress or minimize their emotional reactions due to gender-specific expectations. \cite{vogel_boys_2011, wolter_aufwachsen_2020} This could result in them internalizing the negative aspects of teamwork more strongly and evaluating it as less positive. Comparable correlations cannot be demonstrated for women, at least on the basis of the available data. Nevertheless, potential influences on TWQ can already be identified, such as for H, E, A or O. The contrasting influence of agreeableness between men and women should be emphasized - while men show a positive influence in A, it is negative for women. The sample study suggests that women may be more conflict averse than men, tending to avoid conflict or believing that it negatively impacts team effort and cohesion. Another hypothesis is that women have a greater need for harmony, meaning that any conflicts that do occur have a more significant negative impact on perceptions of COMT and COHS. \cite{wolak_conflict_2022, feng_effects_2023}

The correlations at the indiv-team level could only be determined for the total sample and the men. For the men, team duration, i.e., how long the person has been a member of the respective team, has a negative correlation with honesty and conscientiousness. It is possible that individuals with higher H and C tend to be part of a team less often or change teams more often. TeamDuration and TeamHours are positively related to some TWQ facets, especially CMNC. Apparently, teamwork is perceived more positively the longer members work together.

Individual O is negatively related to the averaged TWQ facets. At the same time, individual O is not significantly related to the unaveraged TWQ facets. Open men may tend to either work independently or express their opinions freely. This could lead them to integrate less into teamwork or cause more conflict. This behavior would also be reflected in the average TWQ score. \cite{feingold_gender_1994} found that men are more assertive and less fearful, suggesting that outspoken men are indeed more inclined to express their opinions openly and thus may create conflict.

Interestingly, individual O also correlates with the average age of team members, although individual O does not correlate with individual age. Average age, in turn, correlates negatively with TWQ facets. It is possible that there are age-related differences in work and communication behavior among the men studied. Thus, it would be possible that older team members act more conservatively, while younger team members are more open and accordingly prefer other approaches. Different communication behavior has already been demonstrated in studies: Younger people, for example, have less difficulty communicating in a way that is appropriate for the addressee or adapting the communication style to the interlocutor \cite{horton_age-related_2007}.
In the overall sample, the correlations turn out to be less pronounced than for men. Because the results at the indiv-indiv level vary by gender, it is possible that different correlations exist for women than for men. Therefore, the correlations in the total sample, which is predominantly men, may be less significant. This is similar at the team-team level.

At the male team-team level, several standard deviations of the HEXACO dimensions correlate negatively with TWQ. Accordingly, personality homogeneous teams seem to have a higher TWQ, with homogeneity seeming to be most important for E, A, and H. It is possible that value conflicts occur when values in the above-mentioned HEXACO dimensions deviate strongly from each other. At the same time, average and standard deviation of age correlate negatively with TWQ facets. The negative correlation with standard deviation could indicate that peers get along better with each other.
A possible rationale for the negative correlation with the age average would be the following: Since no significant negative correlation between age and the TWQ facets could be demonstrated at the Indiv-Indiv level, it cannot be assumed that older team members evaluate TWQ more critically. Instead, the figures in this study suggest that TWQ is more positive in young teams. Possible reasons for this are similar educational backgrounds, lower attachment to established practices, or high motivation. Higher motivation has been shown in other studies \cite{hertel_age_2013}. It would also be possible that older teams communicate less and tend to avoid conflict, which several studies suggest \cite{zhaoyang_age_2018, luong_better_2011}. However, it must be noted that the students studied have a low age range. In this respect, the correlation may hold for the age of students, but it does not necessarily hold for all age groups. It is also conceivable that students lose motivation as the number of semesters increases, and thus as their age increases. A follow-up study could elucidate whether the negative age correlation applies only to undergraduates or to all age groups.

These results may provide part of the answer to the still controversial question of whether diversity in teams has a positive or negative impact \cite{zahl_teamwork_2023}. While professional diversity seems to have a positive impact, character diversity and age diversity have a negative impact according to these results.
Also, at the team-team level, the average O score of men correlates negatively with TWQ and TWQ facets. This correlation is also the only significant correlation between TWQ and the HEXACO dimensions for men. 
The male sample and overall sample have comparable correlation coefficients for many variables. However, two variables show differences between genders: In male software teams, a high diversity of X may have a positive influence on TWQ, although the standard deviation of X currently has a non-significant, partly negative influence on some TWQ facets in the total sample. It is unclear whether this trend holds for women or mixed teams. Additionally, there is a gender-specific difference in the average H score. Although men may have a negative influence, the overall sample shows a weakly positive influence. It is possible that women have a positive relationship between their average H score and the TWQ facets. 
Currently, the proportion of women has a positive impact on TWQ, but it is not yet statistically significant. In summary, various factors are conducive to TWQ, although gender-related differences should certainly be taken into account. The results of this study were able to demonstrate that certain personality traits, but also their composition within teams, influence TWQ. In addition, average age and age diversity also have a non-negligible influence on the TWQ result.

\subsection{Hypothesis testing \& Answering the RQ}

The correlation analysis between HEXACO scores and TWQ ratings revealed that individuals with high X, high A, high C, and low E rate TWQ higher. Thus, H and O have no significant effect on TWQ scores. When these results are separated by gender, it appears that for male participants, C has no significant influence. For female participants, there are indications that significantly different personality dimensions have an influence. For example, A has a negative influence in contrast to the men. 
There are also indications that H and O have an influence, while X has less influence. Hypothesis H1 can therefore be confirmed. In the area of female teams, however, further studies are necessary to verify the recorded figures and to add further significances.
While only a few essential correlations were demonstrated in the overall sample, a negative correlation exists between individual O as well as the TWQ mean score in the team among male respondents.

The total sample does not confirm Hypothesis H2. However, it is true for men, as a relationship has been demonstrated here. For the women, no investigation was possible, since too few teams with a sufficient number of female members were available.
H3.1 can be confirmed, as different personality dimensions have a positive influence on TWQ and its facets. For example, in the team, X has a very strong influence on COMT or dedication, and C has a strong influence on COHS. For men, on the other hand, O has a very negative influence on TWQ and its facets. H3.2 can also be confirmed. The results of this study show that high diversity in personality dimensions has a partly positive and partly negative influence on TWQ facets. For example, BLNC is strongly influenced by the standard deviations of E and O. Particularly in male teams, diversity is shown to have a negative influence on all dimensions except X.
In addition to the hypotheses formulated, it was also possible to demonstrate that age, duration of membership in the team and time spent working in the team also have an influence. According to this, young teams with a low age diversity have the best TWQ. The TWQ also increases the longer the team members have been part of the team and the more working hours they invest in teamwork per week. This answers the research question about the influence of personality traits on TWQ as follows: Personality traits have a strong influence on TWQ. A distinction must be made between same-gender and mixed-gender teams, as the personality traits influencing TWQ vary by gender. The composition of teams in terms of diversity of personality traits also has a strong influence on TWQ. To achieve high TWQ scores, certain personality traits should be represented homogeneously and others inhomogeneously in the team.

\subsection{Limitations}

Despite the promising initial results, the following limitations must be discussed for the study design used. It is assumed that the dimensions of BFI can change considerably during childhood and adolescence and only stabilize to a large extent after the age of 30 \cite{specht_stability_2011}. Since HEXACO has its origins in BFI, it can be assumed that HEXACO dimensions may also vary. This would also fit with the influencing factors of personality, since processes such as cognitive development are strongly pronounced in childhood and adolescence. Thus, it can be implied that the participants in this study may also have personality development that has not yet been completed and thus could not provide a self-assessment that is stable over time. This complicates the generalizability of the present results. 

There are only a few studies on TWQ. This means that there is a lack of sufficient studies to validate TWQ. Additionally, different models or operationalizations of TWQ exist, which differ significantly from each other and only have limited correlation. \cite{silva_comparative_2021} For this reason, it is not known whether the choice of an alternative operationalization of the TWQ would have led to better results.

Several factors must be considered when assessing the representativeness of this sample. First, it should be noted that the selection of teams in the non-probability sample may have been influenced by their proximity to University of Applied Sciences Aachen. This could lead to bias, as teams at other universities may have different personality traits and work styles. Selection bias cannot be ruled out within the university either, as the teams do not cover all semesters. The women share of 22~\% corresponds to the women share of computer science students in Germany \cite{komm_mach_mint_datentool_2020} - however, it was not possible to determine the proportion of women in student software teams. Furthermore, the number of teams studied is relatively small, which could limit the generalizability of the results. A probability sample with a larger sample size could address the issue of selection bias. Additionally, the sample's unbalanced gender distribution could introduce bias, requiring further studies to verify any statements about gender-specific differences. Alternatively, gender could be used as a covariate instead of a group variable to reduce statistical bias.
Study participants did not participate in the survey under laboratory conditions. Instead, they participated at random times in unrecorded locations. Trait-theoretic personality models are based on the basic assumption that personality is stable over time and across situations, practice shows that the self-descriptions used in the models can vary depending on the situation. For example, athletes describe themselves as more psychologically stable when interviewed in the context of their sport. Outside of this, the results turn out to be much worse. The tendency to present oneself in a certain light can also have an influence on the answers to the questions and thus on the quality of the survey. \cite{becker_grundlagen_2021} An analogy may have occurred with the students interviewed in this paper, as they may have been biased by the university environment or professors present. The \textit{Ease of retrieval} effect, which states that agreement with certain attributes becomes more likely when information about them is more easily retrieved from memory, would also support this possible circumstance. \cite{fischer_sozialpsychologie_2018} The sampling method is also susceptible to bias due to acquisition via the team leader, as the selection of participants could be influenced by the team leader. This could be prevented by contacting teams directly.
As already presented in section \ref{sec_background}, teams go through different development phases according to \cite{tuckman_stages_1977}. In the data collection conducted, it was not queried/reviewed which phase the teams were in. While it was recorded how long a team member has been on a team, team development phases can vary in length. In addition, any change in personnel can affect progress in the development phases. Thus, it is difficult to classify the TWQ values recorded. Future research could start here and observe the TWQ of newly formed teams over a longer period of time in order to test the above assumption. If the assumption proves true, identifying trends in a team's TWQ could allow early detection of a move to a different team development phase. This would allow teams to be better supported in the appropriate phases. Furthermore, this would allow a statement to be made as to whether teams with certain personality compositions complete individual phases better/faster.

\section{Conclusion}

The aim of this study was to develop a study design answering the question what influence personality traits have on teamwork quality (TWQ) in student software teams.
For this purpose, a preliminary primary data collection ($n=54$) was conducted, in which the HEXACO personality as well as the TWQ were recorded by means of a digital questionnaire. The collected data were then analyzed in the context of a correlation analysis. The analysis showed that personality traits and their composition significantly influence TWQ. Gender differentiation was informative, as exclusively male teams were influenced by different personality traits compared to female and mixed-gender teams. Other variables, such as the percentage of women and age distribution, also impacted TWQ.
The study's initial results demonstrate the study design's usefulness and validity. Repetition of the presented methodology could yield promising results for practice and research, considering the mentioned limitations.

\section*{Supplementary Material} \label{sec_suppleMat}
Please visit the following URL for raw data, tables, survey questions, and additional materials: \url{https://doi.org/10.17605/OSF.IO/JB94W  }

\section*{Further Notes}
The study was conducted in 2023 and is published here retrospectively.

\bibliographystyle{IEEEtran}
\bibliography{literature}

\end{document}